%% file: gIcoa_astroph.tex
\documentclass[mypaper,7pt,twoside]{CoAst}
\usepackage{epsf,graphicx,fancyhdr}
\usepackage{natbib}
\input{CoAst_layo}

\newcommand{\be}{\begin{equation}}
\newcommand{\bea}{\begin{eqnarray}}
\newcommand{\ee}{\end{equation}}
\newcommand{\bc}{\begin{center}}
\newcommand{\ec}{\end{center}}
\newcommand{\bt}{\begin{table}}
\newcommand{\et}{\end{table}}
\newcommand{\eea}{\end{eqnarray}}
\newcommand{\bfig}{\begin{figure}}
\newcommand{\efig}{\end{figure}}
\newcommand{\bfige}{\begin{figure*}}
\newcommand{\efige}{\end{figure*}}

\def\dss {$\delta$ Scuti stars}

\def\teff {{T}_{\mathrm{eff}}}

\def\vsini {{\vr\!\sin\!i}}
\def\vr {{V}}

\def\BU {{BU\,Cnc}}
\def\BN {{BN\,Cnc}}
\def\BW {{BW\,Cnc}}
\def\BS {{BS\,Cnc}}
\def\BV {{BV\,Cnc}}

\def\anl {{\alpha_{\mathrm{NL}}}}
\def\amlt {{\alpha_{\mathrm{MLT}}}}

\def\michel {{M99}}

\input abrev.tex
\begin{document}
\sf

\chapterDSSN{Asteroseismology of \dss\ in open clusters: Praesepe}
{J.C. Su\'arez, E. Michel, G. Houdek, F. P\'erez Hern\'andez, Y. Lebreton}

\Authors{J.~C. Su\'arez$^{1,2}$, E. Michel$^{2}$, G. Houdek$^3$, 
         F. P\'erez Hern\'andez$^{4,5}$, Y. Lebreton$^6$} 
\Address{$^1$ Instituto de Astrofísica de Andalucía (CSIC), CP3004, Granada, Spain
         $^2$ LESIA, Observatoire de Paris-Meudon, UMR8109, France\\
         $^3$ Institute of Astronomy, University of Cambridge, Cambridge CB30HA, UK\\
	 $^{4}$Instituto de Astrof\'{\i}sica de Canarias (IAC), Tenerife, Spain \\          
         $^{5}$Departamento de Astrof\'{\i}sica, Universidad de La Laguna, Tenerife, Spain\\
         $^{6}$GEPI, Observatoire de Paris-Meudon, Meudon, France}

\noindent
\begin{abstract}
      The present paper provides a general overview of the asteroseismic
      potential of \dss\ in clusters, in particular focusing on convection
      diagnostics. We give a summarise of the last results obtained by
      the authors for the Praesepe cluster of which five \dss\ are 
      analysed. In that work, linear analysis is confronted with
      observations, using refined descriptions for the 
      effects of rotation on the determination of the global stellar 
      parameters and on the adiabatic oscillation frequency computations.
      A single, complete, and coherent solution for all the selected
      stars is found, which lead the authors to find important restrictions
      to the convection description for a certain range of effective
      temperatures. Furthermore, the method used allowed to give an estimate
      of the global parameters of the selected stars and constrain the
      cluster. 
\end{abstract}


\section{Introduction\label{sec:intro}}

The main idea of the work here outlined is to compare ranges of observed and predicted 
unstable modes and analyse them in terms of ranges of radial modes. 
That methodology was conceived and used in \citet{MiHer99}, hereafter \michel,
and here is revisited. The use of radial modes is justified because 
driving and damping in \dss\ takes place predominantly in the He\,{\sc ii} ionisation 
zone, which is rather close to the stellar surface, where the vertical scale is much 
less than the horizontal scale of the oscillations and when $\ell$ is low the modal inertia
is quite insensitive to degree $\ell$. We consider the following stars belonging to the 
Praesepe cluster: \BW, \BS, \BU\, and \BN\  (already included in the sample considered 
by \michel), which were observed by several campaigns of the STEPHI network \citep{Michel95}, 
and a fifth star, \BV, which was observed by \citet{Frandsen01}.

The present work intends to search for one particular solution 
that explains the whole set of observations, instead of sets of individual solutions 
for each star (methodology followed by \michel\ whose results are considered here
as the reference domain of possible solutions). To do so, refined techniques for modelling 
intermediate mass stars are required, in
particular those taking different effects of rotation into account. 
This requirement is fulfilled in this work with modern techniques that
largerly improve those adopted in \michel\ and other precedent works.
One of the major improvements is the use the complete second-order formalism (including near
degeneracy effects) developed by \citet{Sua06rotcel} based on the works
by \citet{DG92} and \citet{Soufi98}, which, in addition,
takes the effect of the star deformation due to rotation into account.

\section{Procedure\label{sec:procedure}}

Firstly, we calculate the fundamental parameters of each selected star.
To do so we apply the method by \citet{Pe99} which corrects the photometric observables
for the effect of rotation. That method is model-dependent, so that solutions are
obtained by adjusting some typically free parameters. The best solution 
given by the correction for the effect of rotation was obtained for 
$\amlt=1.614$ and $d_\mathrm{ov}=0.2$. This best solution correspond to 
an age of the cluster of 650~Myr ($\pm20-40$ Myr). The age uncertainty of 20--40 Myr can be neglected
in terms of global characteristics of the non-rotating co-partners
\citep[see the influence of age on photometric corrections for rotation
in different open clusters in][]{Sua02aa}. 

The fundamental parameters obtained are then used to build representative asteroseismic models 
for each star, consisting of pseudo-rotating equilibrium models and their corresponding 
adiabatic oscillation spectra. Theoretical adiabatic oscillation spectra are computed 
with the oscillation code {\sc filou} \citep{filou,SuaThesis} which uses the complete 
treatment of second-order effects of rotation by \citet{Sua06rotcel}, based on the 
formalisms by \citet{DG92} and \citet{Soufi98}. These asteroseismic models
are then used to determine the \emph{observed} radial orders ranges, which,
as it was done in \michel, are then confronted with mode instability predictions obtained
from a linear stability analysis. 

Unstable radial modes are calculated by means of linear stability computations
which are carried out in the manner of \citet{Balmforth92}. 
We adopt the nomenclature of \michel, i.e. $\alpha_{\rm NL}=l/H_{\rm p}$ 
is the mixing-length parameter of the non-local convection model used in the 
stability computations. The non-local mixing-length parameter $\alpha_{\rm NL}$
is calibrated to the same depth of the outer convection zone as suggested 
by the evolutionary computations which use the standard mixing-length 
formulation by \citet{BohmVit58} and a local mixing-length parameter 
$\alpha_{\rm MLT}=1.614$, for which the equivalent calibrated value of 
$\alpha_{\rm NL}=1.89$ is obtained. As in \michel, a second series
of stellar models with $\alpha_{\rm NL}=1.50$ in order to analyse the
impact of varying the mixing-length on mode stability.
\begin{figure}
  \centering
  \includegraphics[width=8cm]{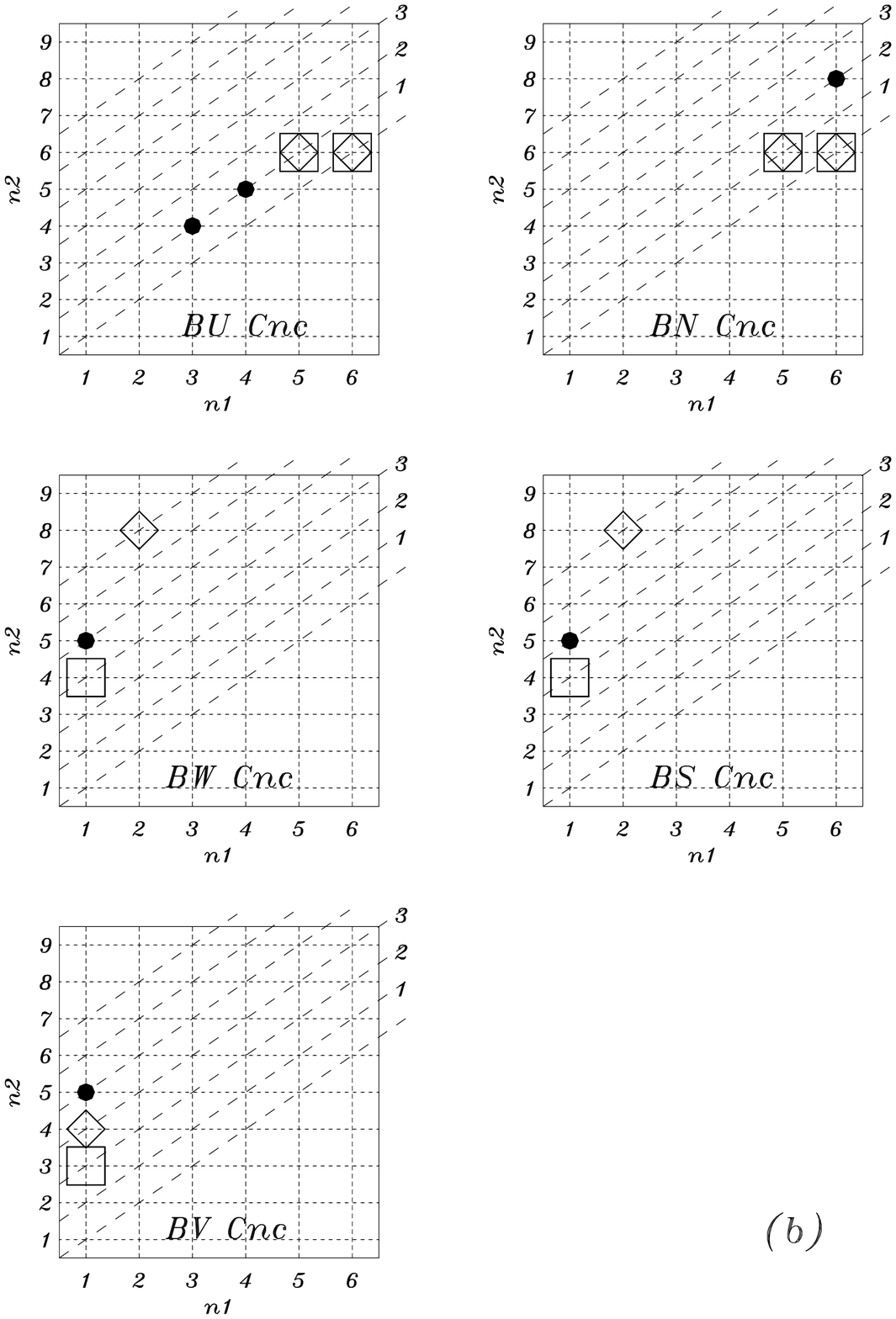}
  \caption{Observed and predicted (using linear stability analysis) ranges of unstable radial 
           modes for the selected \dss\ ($n_1$ is the lowest value, and
           $n_2$ the largest value of the radial order of the unstable modes)
	   displayed for the selected \dss. Filled circles represent the observed ranges. 
	   Rhombus and squares correspond to predicted radial order ranges for $\anl=1.89$ and
	   $\anl=1.50$, respectively. Each diagonal-dashed line represents the width (in
	   radial orders) of the represented ranges. Taken from \citet{Sua07gammes}.}
  \label{fig:GammesPrae}
\end{figure}
%

\section{Comparison between observed and predicted ranges of unstable 
         radial modes\label{sec:Compar}}

We compare ranges of observed and predicted
radial orders $n$ of unstable modes for two values of $\anl$:
1.89 and 1.50. This comparison is shown in Fig.~\ref{fig:GammesPrae} in
which an uncertainty of $\pm1\,n$ in the determination of the range of radial orders
\citep[see][for a detailed explanation of uncertainty sources]{Sua07gammes}.
For each star, we have two representative models 
correspondig to considering $\pm10\%$ of the observed $\vsini$ value, so we
calculate four radial order ranges are calculated per object. 
It is found that, in general, the observed ranges are in good agreement with 
the theoretical predictions using $\anl=1.50$, whereas with $\anl=1.89$ the predicted 
ranges are in grave disagreement with the observations for the intermediate mass stars 
\BW\ and \BS, which present a temperature range of $\log\,\teff=3.87$\,--\,$3.88$ .
For these stars a smaller mixing-length parameter $\anl$ is then required than suggested 
from a calibrated solar model. The need of a smaller value for $\anl$
than that from a calibrated solar model was also reported by
\citet{Pagoda05} for the $\delta$ Scuti star FG Vir.
For the massive objects, \BU\ and \BN, the predicted unstable ranges are 
compatible with the observed results ($\pm1\,n$), for both $\anl=1.50$ and
$\anl=1.89$. The results for both objects are thus not sensitive to the value of $\anl$.
This is to be expected because these more massive stars have shallower 
outer convection zones and thus their structures are less sensitive to the assumed value 
of the mixing-length parameter. Finally, for the less massive object, \BV, the values for 
$\anl$ cannot be distinguished. Nevertheless, the observed ranges agree with the theoretical
predictions within $\pm1\,n$. Therefore, in general, these results constitute a consistent solution
in terms of physics and cluster membership, and the observed and theoretical ranges of radial orders 
are in reasonable agreement for all the stars considered in this work. 

Details of the procedure, model characteristics, as well as the detailed comparison
between the present results with those of \michel\ are given in a forthcoming
paper \citep{Sua07gammes}. We note that, as in \michel, the instability predictions
are carried out using equivalent envelope models which do, however, not take the effect of
rotation into account. This is so because, up to date, there are no reliable theories 
available which describe the effect of rotation on mode stability. Nevertheless, 
in a forthcoming paper \citep{Sua07gammes}, a crude estimate of this effect is addressed, 
which assumes that mode stability depends predominantly on the effective temperature of the 
model \citep{Alosha75}. As well, in that work, more details on the procedure, the
characteristics of models, and a detailed comparison between the present results with those of 
\michel\ are provided.

\acknowledgments{JCS acknowledges support by the Instituto de 
Astrof\'{\i}sica de Andaluc\'{\i}a by an I3P contract
financed by the European Social Fund and from the Spanish 
Plan Nacional del Espacio under project ESP2004-03855-C03-01.
GH acknowledges support by the Particle Physics and Astronomy
Research Council of the UK. }

\bibliographystyle{aa}
\bibliography{/home/jcsuarez/Boulot/Latex/Util/References/ref-generale}

\end{document}

%% file: CoAst_layo.tex
\renewcommand{\chaptermark}[1]{\markboth{#1}{}}

\newcommand{\dss}{$\delta$~Scuti stars}                     

\newcommand{\teff}{\ensuremath{T_{eff}}}             
\newcommand{\vsini}{\ensuremath{v\sin i}}                   

\newcommand{\kopf}{\small\itshape Comm. in Asteroseismology\\ Vol. number, publication date (will be inserted in the production process)}
\newcommand{\Authors}[1]{\begin{center}\normalsize\bf\sf #1 \end{center}}

\renewcommand{\author}[1]{\begin{center}\normalsize\bf\sf #1 \end{center}}
\newcommand{\Address}[1]{\begin{center}\small\sf #1 \end{center}}

\renewenvironment{abstract}{\section*{Abstract}\normalsize\sf}{}

\newcommand{\chapterDSSN}[2]{\chapter[\sf\normalsize #1\\ \footnotesize \hspace*{5mm}by #2 \sf\normalsize][]{#1\\}\rhead[\fancyplain{}{\sf\footnotesize \center{#1}}]{\fancyplain{}{\sffamily\thepage}}\lhead[\fancyplain{\kopf}{\sffamily\thepage}]{\fancyplain{\kopf}{\sf\footnotesize \center{#2}}}}

\renewcommand{\chaptername}{}
\newcommand{\acknowledgments}[1]{\vspace*{5mm}\noindent\begin{bf}Acknowledgments. \end{bf} #1}